\documentclass[twocolumn,aps,superscriptaddress,multicol,amsmath,amssymb]{revtex4-2}

\makeatletter
\newcommand*{\rom}[1]{\expandafter\@slowromancap\romannumeral #1@}
\makeatother
\usepackage{xcolor}
\usepackage{color}
\usepackage{graphicx}
\usepackage[colorlinks=true, linkcolor=blue, urlcolor=blue, citecolor=blue]{hyperref}
\usepackage{amsmath,amssymb}
\usepackage{epstopdf}
\usepackage[font=small,labelfont=bf,justification=justified]{caption}
\usepackage[font=small,labelfont=bf,justification=justified]{subcaption}

\usepackage{amssymb}
\usepackage{amsmath}
\usepackage{graphicx}
\usepackage{epstopdf}
\usepackage{bm}
\usepackage{gensymb}
\usepackage{soul}
\usepackage{sidecap}
\usepackage[normalem]{ulem}
\usepackage{times}
\usepackage{float}
\usepackage{enumerate}
\usepackage{multirow}
\usepackage{tabularx}
\usepackage{array}
\usepackage{url}
\usepackage{slantsc}
\usepackage{lmodern}
\usepackage[normalem]{ulem}
\usepackage{xparse}
\usepackage{soul}
\usepackage{xcolor}
\makeatletter
\NewDocumentCommand{\sotwo}{O{red}O{black}+m}
{%
	\begingroup
	\setulcolor{#1}%
	\setul{-.5ex}{.4pt}%
	\def\SOUL@uleverysyllable{%
		\rlap{%
			\color{#2}\the\SOUL@syllable
			\SOUL@setkern\SOUL@charkern}%
		\SOUL@ulunderline{%
			\phantom{\the\SOUL@syllable}}%
	}%
	\ul{#3}%
	\endgroup
}
\makeatother
\graphicspath{{figs/}}
\def\beq{\begin{equation}}
\def\eeq{\end{equation}}
\def\bea{\begin{eqnarray}}
\def\eea{\end{eqnarray}}

\begin{document}

		\title{Thermodynamic Geometry of an Ideal Quon Gas}
	\author {Leila Yousefzadeh}
	\email{leila.yousefzade@gmail.com}
	\affiliation{Department of Physics, University of Mohaghegh Ardabili, P.O. Box 179, Ardabil, Iran}
	\author {Hosein Mohammadzadeh}
	\email{mohammadzadeh@uma.ac.ir}
	\affiliation{Department of Physics, University of Mohaghegh Ardabili, P.O. Box 179, Ardabil, Iran}
	\author {Habib Esmaili}
	\email{habibsmayli@gmail.com}
	\affiliation{Department of Physics, University of Mohaghegh Ardabili, P.O. Box 179, Ardabil, Iran}
		\author {Zahra Ebadi}
	\email{z.ebadi@uma.ac.ir}
	\affiliation{Department of Physics, University of Mohaghegh Ardabili, P.O. Box 179, Ardabil, Iran}
		\author {Morteza Nattagh Najafi}
    \email{morteza.nattagh@gmail.com }
	\affiliation{Department of Physics, University of Mohaghegh Ardabili, P.O. Box 179, Ardabil, Iran}


\begin{abstract}
We investigate the equilibrium thermodynamics and {\color{black}thermodynamic geometry} of an ideal quon gas within the grand canonical ensemble. By incorporating the algebraic deformation parameter $q$, the system generalizes standard bosonic behavior while recovering the conventional ideal Bose gas in the undeformed limit. A rigorous examination of the ground-state occupation reveals two disconnected mathematical domains of the fugacity. By enforcing thermodynamic continuity, single-valuedness, and consistency with the high-temperature classical limit,  we exclude the second mathematical branch and establish the lower interval as the unique physically admissible state space.
	 This identifies the deformation parameter as an intrinsic, generalized Bose--Einstein condensation threshold. Employing the Fisher-Rao metric on the equilibrium parameter manifold, we probe the thermodynamic scalar curvature across all temperature regimes. The scalar curvature remains strictly positive throughout the physical domain, confirming that the deformation preserves an effectively attractive statistical interaction without inducing fermionic tendencies. Near the critical condensation threshold, the curvature increases sharply and exhibits a {\color{black} divergence}, providing an unambiguous geometric signature of macroscopic coherence and critical fluctuations. Below the transition temperature, the pinning of fugacity eliminates a fluctuating degree of freedom, collapsing the scalar curvature to zero.
\end{abstract}

\maketitle
	
\section{Introduction}\label{1}
Statistical mechanics provides the foundational bridge connecting the microscopic laws of quantum theory with the macroscopic phenomenology of many-body matter~\cite{pathria1996statistical}. In standard quantum mechanics, the permutation symmetry of identical particles leads inexorably to the dichotomy of Bose--Einstein and Fermi--Dirac statistics~\cite{hu1988progress}. This conventional framework has achieved {\color{black}considerable success} in capturing a vast spectrum of physical behaviors, ranging from the macroscopic coherence of Bose--Einstein condensates to the structural stability of degenerate electron gases in condensed matter and astrophysics~\cite{hutchinson1997finite,mahan2000introductory}. 
Nevertheless, continuous frontiers in strongly correlated quantum matter, low-dimensional systems, and quantum gravity have revealed scenarios where this dichotomy proves restrictive~\cite{haldane1991fractional,Powell02042020,paschen2021quantum,morel2022fractionalization,dhar2025observing}. Consequently, generalized quantum statistics have emerged as a fertile domain of investigation, seeking to interpolate smoothly between bosonic and fermionic limits or to describe novel statistical behavior that departs fundamentally from conventional exchange paradigms~\cite{fivel1990interpolation,hasegawa2009bose}.

Among various generalized schemes, the quon algebra formulated by Greenberg stands out as a mathematically rigorous and structurally elegant approach~\cite{greenberg1990example}. The quon framework introduces a continuous deformation parameter $q \in [-1, 1]$ into the commutation relations of creation and annihilation operators while preserving a strictly positive-definite Fock space norm~\cite{chaichian1993statistics}. Unlike certain fractional statistics that are strictly confined to two spatial dimensions, quons can be formulated consistently in arbitrary spatial dimensions. The parameter $q$ interpolates between the symmetric Bose algebra at $q=1$ and the anti-symmetric Fermi algebra at $q=-1$, while intermediate values give rise to deformed quantum states with intriguing statistical correlations. In thermodynamic settings, studying how this deformation alters the grand canonical potential, the density of states, and the onset of macroscopic condensation offers valuable insights into the fundamental role of particle identity in quantum systems~\cite{ciolli2022thermodynamics}.

Parallel to algebraic deformations, the geometric perspective of equilibrium thermodynamics has evolved into a powerful diagnostic tool~\cite{ruppeiner1995riemannian}. Originating from the seminal work of Weinhold and subsequently generalized by Ruppeiner, thermodynamic geometry equips the manifold of equilibrium states with an intrinsic Riemannian metric derived from fluctuation theory and the Fisher--Rao information metric.
{\color{black}
For the interpretation of thermodynamic curvature and a broad overview of thermodynamic geometry, see Refs.~\cite{Ruppeiner2010,Mahmoudi2023}; its application to ideal quantum gases with a relativistic dispersion is studied in Ref.~\cite{Mohammadzadeh2026}.}
 Within this framework, the thermodynamic scalar curvature $R$ carries profound physical meaning: its magnitude measures the effective correlation volume and the intensity of microscopic fluctuations, while its divergence serves as an invariant hallmark of continuous phase transitions and critical phenomena~\cite{ruppeiner1995riemannian,rodrigo2025interacting}. Furthermore, the empirical sign of $R$ has been widely recognized as an indicator of effective statistical interactions, displaying negative values for effective statistical repulsion (as in fermion gases) and positive values for effective statistical attraction (as in boson gases)~\cite{mirza2009nonperturbative}. 
\textcolor{black}{Following the Ruppeiner framework, $R$ provides a geometric measure of the effective statistical correlations and interactions encoded in the equilibrium fluctuations of the system.Since the Ricci scalar is invariant under coordinate transformations, the results are independent of the particular representation chosen for the Fisher--Rao metric~\cite{ruppeiner1995riemannian}.}
Although extensively applied to ideal and interacting classical gases, standard quantum gases, magnetic models, and black hole thermodynamics, the geometric properties of deformed quantum statistics—and quons in particular—remain largely unexplored. Important open questions include how the deformation parameter reshapes the manifold of states, alters the boundaries of the physically accessible thermodynamic domain, and manifests near condensation thresholds.

In this work, we present a systematic investigation of the thermodynamics and thermodynamic geometry of an ideal quon gas in the grand canonical ensemble. We demonstrate that the algebraic requirement of ground-state occupancy non-negativity divides the fugacity into distinct domains.
 \textcolor{black} {By imposing thermodynamic continuity,
	single-valuedness, and consistency with the high-temperature classical limit, we
	identify the physically admissible branch of the present model as $0 < z \leq q$}.
	 This identifies the deformation parameter itself as the critical condensation threshold, $z_c = q$. By constructing the Fisher--Rao metric, we obtain exact analytical closed-form expressions for the thermodynamic scalar curvature in terms of generalized polylogarithms and study its behavior across all temperature regimes. We demonstrate that the scalar curvature diverges at the critical temperature, providing a {\color{black} geometric signature} of condensation, and collapses to zero in the condensed phase due to the dimensional reduction of the fluctuating state space.

The remainder of this paper is structured as follows. 

{\color{black}
In Sec.~\ref{2}, we formulate the statistical framework of the quon gas, introduce the
grand canonical partition function by adopting the single-mode result established in
Ref.~\cite{chung2026quon}, and determine the
admissible domain of the fugacity from the ground-state occupancy condition.}
 In Sec.~\ref{3}, we obtain the fundamental thermodynamic potentials, equation of state, and response functions. In Sec.~\ref{4}, we review the Riemannian formulation of thermodynamic geometry based on the Fisher--Rao metric and fluctuation theory. In Sec.~\ref{5}, we apply this geometric framework to the quon gas, deriving explicit expressions for the metric tensor and the thermodynamic scalar curvature. In Sec.~\ref{6}, we resolve the physical interpretation of the fugacity branches, analyze the temperature dependence of the chemical potential, and uncover the geometric behavior of the curvature across both the normal and condensed phases. Finally, Sec.~\ref{7} presents our concluding remarks and discusses promising avenues for future research.

\section{Statistical and Thermodynamic Properties of the Quon Gas}\label{2}
While standard Bose--Einstein and Fermi--Dirac statistics provide an exact description of non-interacting quantum gases composed of fundamental particles, this conventional framework encounters limitations when collective excitations display statistical characteristics that deviate from strict permutation symmetries~\cite{haldane1991fractional,fivel1990interpolation,greenberg1990example}. In strongly correlated quantum matter, macroscopic properties are often dominated by emergent quasiparticles rather than the underlying bare constituents~\cite{mahan2000introductory}. This realization has stimulated significant interest in generalized quantum statistics built upon deformed oscillator algebras, establishing a rigorous framework to describe complex excitations such as excitons, magnons, and polarons that depart from canonical exchange symmetries~\cite{arik1976hilbert,biedenharn1989quantum,macfarlane1989ionmix}. These deformed algebras generalize the standard Heisenberg commutation relations through continuous deformation parameters and maintain deep algebraic connections to quantum groups and non-commutative geometry~\cite{biedenharn1989quantum,macfarlane1989ionmix}. Furthermore, while fractional braid statistics such as anyons are strictly topologically confined to two spatial dimensions, deformed-algebra statistics offer the conceptual advantage of being well-defined in arbitrary spatial dimensions~\cite{wilczek1982quantum,greenberg1990example}.

Among the diverse formulations of generalized quantum statistics, the quon algebra introduced by Greenberg provides a particularly transparent and analytically tractable framework that continuously interpolates between bosonic and fermionic statistics~\cite{greenberg1990example}.
 In contrast to generic $q$-deformed oscillators where deformation typically alters single-mode number operators nonlinearly, the quon framework introduces the deformation directly into the bilinear exchange relations while preserving a positive-definite Fock space representation~\cite{greenberg1990example,greenberg1991particles,PhysRevE.61.1218}.

The fundamental quon algebra is governed by the minimally deformed commutation relation~\cite{greenberg1990example,greenberg1991particles}
\begin{equation}\label{Eq:algebra}
	\begin{aligned}
		& a^{}_{i} a^{\dagger}_{j} - q a^{\dagger}_{j} a^{}_{i} = \delta^{}_{ij}, \\
		& a^{}_{i} \lvert 0 \rangle = 0,
	\end{aligned}
\end{equation}

where $a^{}_{i}$ and $a^{\dagger}_{j}$ represent the quon annihilation and creation operators corresponding to single-particle quantum states $i$ and $j$, $q$ is a real deformation parameter lying in the domain $-1 \le q \le 1$, and $\lvert 0 \rangle$ denotes the unique quonic vacuum state.

Although the quon algebra is formally consistent throughout the full parameter range $-1 \le q \le 1$, the present investigation focuses on the bosonic branch $0 < q \le 1$. 
This regime connects continuously to the ordinary Bose--Einstein limit at $q = 1$, providing a natural arena for exploring Bose-like condensation phenomena and examining how the underlying state-space geometry is reshaped by statistical deformation.
{\color{black}
	We note that the quon gas of Ref.~\cite{chung2026quon} differs from the
	standard quon-gas studies based directly on Greenberg's algebra~\cite{PhysRevA.52.932,PhysRevD.48.2929,GOODISON199438}. In Inomata's
	work~\cite{PhysRevA.52.932}, a specific Hamiltonian,
	$H=\sum_i\epsilon_i(2/[2])^{2N_i-2}a_i^\dagger a_i$, is built on the algebra
	\eqref{Eq:algebra}; the single-level partition function is not available in closed form for
	$-1<q<1$, and the analysis is a fugacity expansion (virial coefficients,
	$d=2$) over $-1\le q\le1$. More generally, such studies found
	thermodynamic properties that are independent of $q$, related to
	Maxwell--Boltzmann statistics, and a lack of extensivity (Gibbs paradox)~\cite{PhysRevD.48.2929,GOODISON199438}.
	 In Ref.~\cite{chung2026quon}, by contrast, the standard
	Hamiltonian is used and the deformation enters through the statistical weight
	(Eq.~(3)~\cite{PhysRevD.48.2929}) and a $q$-twisted KMS condition. This gives closed-form
	$\mathcal{Z}_p$ and $\langle n_p\rangle$ (Eqs.~(5), (8)~\cite{PhysRevD.48.2929}), a thermodynamics
	that depends on $q$, a fugacity domain $0<z\le q$, and a condensation
	transition.}

To establish the equilibrium thermodynamics of the system, we construct the partition function as
\begin{equation}
	\mathcal{Z}(T, V, \mu) = \sum_{\{n_p\}} \mathcal{G}\{n_p\} \exp\left( -\beta \sum_{p} n_p(\omega_p - \mu) \right),
\end{equation}
where $\beta = (k_B T)^{-1}$ represents the inverse temperature, $\omega_p$ denotes the single-particle kinetic energy of mode $p$, and $z = e^{\beta\mu}$ denotes the fugacity associated with the chemical potential $\mu$. 
The summation encompasses all accessible occupation-number configurations $\{n_p\}$, with the statistical weight of each microscopic configuration dictated by the combinatorial factor $\mathcal{G}_{p}(n_p)$. This factor encapsulates the deformed permutation algebra of quons through the relation
\begin{equation}\label{eq:G}
	\mathcal{G}^{}_{p}\{n^{}_p\} = P_{n^{}_p}(x) - P_{n^{}_p-1}(x), \qquad x = \tfrac{1}{2}\left(q + q^{-1}\right),
\end{equation}
where $n^{}_{p}=0,1,2,\cdots$ and $P^{}_n(x)$ represents the Legendre polynomial contribution generated by the underlying quon combinatorial algebra.

A notable feature of this combinatorial weight is that individual momentum modes decouple completely, allowing the grand partition function to factorize into a product over independent single-mode contributions \cite{chung2026quon},
\begin{equation}
	\mathcal{Z}(q) = \prod_p \mathcal{Z}^{}_p(q).
\end{equation}

{\color{black}
The explicit evaluation of the single-mode partition function requires the combinatorial weights $G_p\{n_p\}$ of Eq.~\eqref{eq:G}, which are generated by the quon algebra of Eq.~\eqref{Eq:algebra} and are constructed in detail in Ref.~\cite{chung2026quon}.
Adopting that result, the single-mode partition function takes the closed form}
\begin{equation}\label{Eq:zp}
	\mathcal{Z}^{}_p(q) = \frac{1}{\sqrt{(1 - qze^{-\beta\omega_p})(1 - q^{-1}ze^{-\beta\omega_p})}}.
\end{equation}
Taking the logarithm of this expression and summing over all momentum modes, we obtain the total grand potential logarithm for the ideal quon gas:
\begin{equation} \label{logZ}
	\ln \mathcal{Z}(q) = -\frac{1}{2} \sum_p \left[ \ln(1 - qze^{-\beta\omega_p}) + \ln(1 - q^{-1}ze^{-\beta\omega_p}) \right].
\end{equation}
The total particle number $N$ follows directly from the standard grand canonical relation $N = z(\partial \ln \mathcal{Z} / \partial z) = \beta^{-1}(\partial \ln \mathcal{Z} / \partial\mu)$~\cite{huang1987statistical,pathria2011statistical}. Differentiating Eq.~(\ref{logZ}) yields
\begin{equation}
	N = \frac{1}{2} \sum_p \left[ \frac{1}{q^{-1}z^{-1}e^{\beta\omega_p} - 1} + \frac{1}{qz^{-1}e^{\beta\omega_p} - 1} \right]. \label{eq:N_total}
\end{equation}
Because the total particle number is equivalently defined as the sum over the average mode occupations, $N = \sum_p \langle n_p \rangle$, the average occupation number of a single mode $p$ is directly identified from Eq.~(\ref{eq:N_total}) as
\begin{equation} \label{eq:occupancy}
	\langle n^{}_{p} \rangle = \frac{1}{2} \left[ \frac{1}{q^{-1}z^{-1}e^{\beta\omega_p} - 1} + \frac{1}{qz^{-1}e^{\beta\omega_p} - 1} \right].
\end{equation}
\begin{figure}[t]
	\centering
	\begin{subfigure}[b]{0.8\columnwidth}
		\centering
		\includegraphics[width=\textwidth]{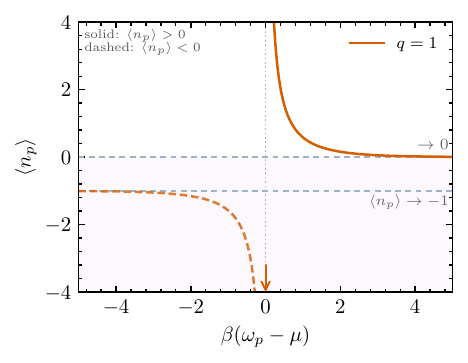}
		\caption{}
		\label{fig:np_a}
	\end{subfigure}
	\hfill
	\begin{subfigure}[b]{0.8\columnwidth}
		\centering
		\includegraphics[width=\textwidth]{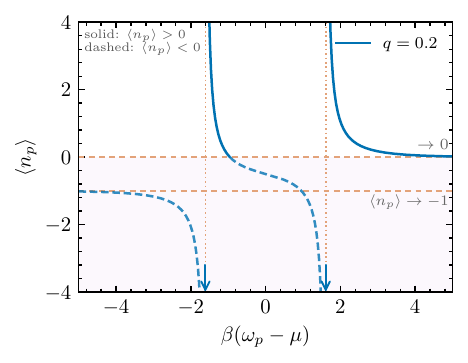}
		\caption{}
		\label{fig:np_b}
	\end{subfigure}
	\caption{\justifying
		Average occupation number $\langle n_p\rangle$ as a function of the reduced parameter $\beta(\omega_p-\mu)$ for representative values of the deformation parameter $q$.}
	\label{fig:n_p_dist}
\end{figure}
Equation~(\ref{eq:occupancy}) reveals that the quon occupation profile can be interpreted physically as the symmetric superposition of two effective Bose-like distributions evaluated at scaled fugacities $qz$ and $q^{-1}z$. In the undeformed limit $q \to 1$, both effective distributions merge smoothly into the standard Bose--Einstein distribution.

Because $\langle n^{}_p \rangle$ represents a physical particle occupancy, it must satisfy the fundamental non-negativity condition $\langle n^{}_p \rangle \ge 0$ for all physical states. For standard ideal Bose gases, this requirement is most restrictive at the single-particle ground state $\omega^{}_p = 0$, where the ground-state occupancy reduces to $\langle n_0 \rangle^{}_B = (z^{-1} - 1)^{-1}$. Demanding $\langle n_0 \rangle^{}_B \ge 0$ yields the familiar upper bound $z \le 1$, which uniquely delineates the allowed fugacity domain~\cite{huang1987statistical,pathria2011statistical}.

In sharp contrast, applying the positivity criterion to the zero-energy ground state of the quon gas introduces qualitative structural differences. Setting $\omega_p = 0$ in Eq.~(\ref{eq:occupancy}) gives
\begin{equation}\label{Eq:nno}
	\langle n_0 \rangle = \frac{1}{2} \left[ \frac{1}{q^{-1}z^{-1} - 1} + \frac{1}{qz^{-1} - 1} \right].
\end{equation}
\begin{figure}[t]
	\centering
	\begin{subfigure}[b]{0.8\columnwidth}
		\centering
		\includegraphics[width=\textwidth]{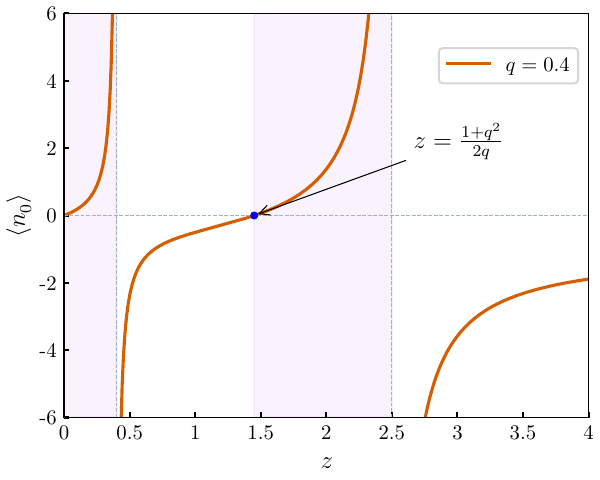}
		\caption{}
		\label{fig:n04}
	\end{subfigure}
	\hfill
	\begin{subfigure}[b]{0.8\columnwidth}
		\centering
		\includegraphics[width=\textwidth]{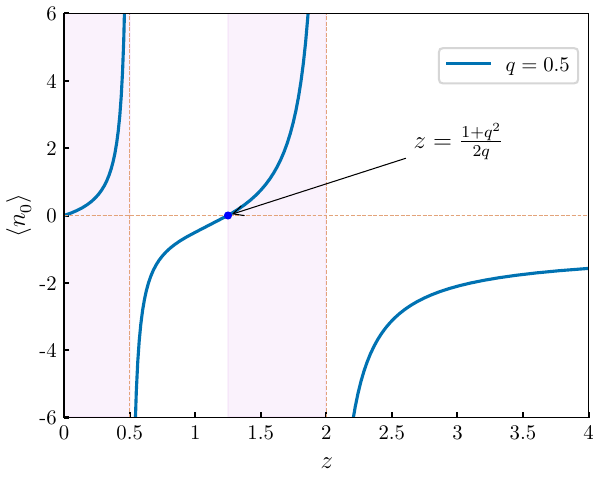}
		\caption{}
		\label{fig:n05}
	\end{subfigure}
	\caption{\justifying
		Zero-energy ground-state occupation number $\langle n_0 \rangle$ as a function of the fugacity $z$ for various values of the deformation parameter $q$. The algebraic poles partition the fugacity domain into disconnected regions. For $0 < q < 1$, the lower interval terminating at $z = q$ constitutes the physically viable thermodynamic state space.}
	\label{fig:n0_z}
\end{figure}

As illustrated in Fig.~\ref{fig:n0_z}, the deformation parameter splits the analytical structure of the occupation function, introducing two distinct algebraic poles at $z = q$ and $z = 1/q$. These singularities partition the fugacity axis into disconnected domains. Consequently, the algebraic requirement $\langle n_0 \rangle \ge 0$ is satisfied across two separate fugacity intervals for $0 < q < 1$:
\begin{align}
	& 0 < z < q,
	\label{part1}\\
	&\textcolor{black} {\frac{1+q^2}{2q}} < z < \frac{1}{q}.
	\label{part2}
\end{align}

These two positive branches are separated by an intermediate gap where the ground-state occupation number becomes negative and hence unphysical. Thus, unlike conventional quantum statistics, non-negativity of the occupancy alone is insufficient to uniquely define the physically realizable state space of the quon gas.

\textcolor{black}{
	As will be discussed in Sec.~\ref{6}, the lower interval $0 < z \le q$ is identified as the physically admissible domain. We argue that incorporating the second interval would lead to an inaccessible temperature range, rendering the system unphysical.}
 Accordingly, the upper bound of this interval defines the deformation-dependent critical fugacity $z_{\mathrm{max}} = z_c = q$. In the following sections, we employ this physically identified domain to construct the thermodynamic potentials, response functions, and Riemannian state-space geometry of the quon gas.

\section{Thermodynamic quantities}\label{3}
Building on the distribution function introduced in the previous section, we now turn to deriving the corresponding thermodynamic quantities of the quon gas.
For this purpose, we consider an ideal gas in $D$ spatial dimensions, characterized by the general energy--momentum dispersion relation \cite{pathria2011statistical}
\begin{equation}
	\epsilon=\alpha p^{\sigma},
\end{equation}
where $p$ denotes the particle momentum, $\sigma$ is the exponent of the dispersion relation, and $\alpha$ is the associated proportionality constant.
This dispersion relation encompasses the two limits of physical interest as special cases: in the non-relativistic limit, where $\epsilon=p^{2}/2m$, one has $\alpha=1/2m$ and $\sigma=2$, with $m$ denoting the particle mass, while in the ultrarelativistic limit, where $\epsilon=pc$, the corresponding parameters are $\alpha=c$ and $\sigma=1$.

Given this dispersion relation, the single-particle density of states in $D$ dimensions takes the form \cite{huang1987statistical,pathria2011statistical}
\begin{equation}
	\Omega(\epsilon)=\frac{A^D}{\Gamma(\frac{D}{\sigma})}\epsilon^{\frac{D}{\sigma}-1},\quad
	A=\frac{L\sqrt{\pi}}{(a^{\frac{1}{\sigma}}h)},
\end{equation}
where $L^{D}$ denotes the volume of the $D$-dimensional box. For simplicity, and without loss of generality, the constant $A$ is set to unity throughout the calculations that follow.

Using the occupation number given in Eq.~(\ref{eq:occupancy}), together with the density of states, the internal energy of the quon gas can now be obtained by integrating over all single-particle energies as
\begin{equation}
	\begin{aligned}
		U=&\int_{0}^{\infty}\epsilon n(\epsilon)\Omega(\epsilon)d\epsilon
		\\=&\frac{\nu}{2} {\beta^{-(\nu+1)}}\left[\text{Li}^{}_{(\nu+1)}(z_{1})+\text{Li}^{}_{(\nu+1)}(z_{2})\right]
		\\=& \nu{\beta^{-(\nu+1)}}\bar{\text{Li}}^{}_{\nu+1}(q,\nu;z),
		\label{internalenergy}
	\end{aligned}
\end{equation}	
where, for notational convenience, we have introduced $\frac{D}{\sigma}\equiv\nu$ together with the shifted fugacities $z_{1}=qz$ and $z_{2}=q^{-1}z$, and defined the generalized polylogarithm function
\begin{equation}		
	\bar{\text{Li}}^{}_{\nu}(q,\nu;z)\equiv\frac{1}{2} \left[\text{Li}^{}_{\nu}(z_{1})+\text{Li}^{}_{\nu}(z_{2})\right],\label{G}
\end{equation}
which reduces to the ordinary Bose--Einstein integral $\mathrm{Li}_{\nu}(z)$ \cite{huang1987statistical,pathria2011statistical}, given explicitly by
\begin{equation}
	\text{Li}_{\nu}(z)=\frac{1}{\Gamma(\nu)}\int_{0}^{\infty}\frac{x^{\nu-1}}{z^{-1}\exp(x)-1}dx,
	\label{polylogarithm}
\end{equation}
in the limit $q\rightarrow1$, where $z_1,z_2\rightarrow z$.

Following the same procedure, the total particle number of the quon gas is obtained as
\begin{equation}\label{eq:N}
	\begin{aligned}
		N=&\int_{0}^{\infty}n(\epsilon)\Omega(\epsilon)d\epsilon
		=\frac{1}{2} {\beta^{-\nu}}\left[\text{Li}^{}_{\nu}(z_{1})+\text{Li}^{}_{\nu}(z_{2})\right]\\&={\beta^{-\nu}}	\bar{\text{Li}}^{}_{\nu}(q,\nu;z).
	\end{aligned}
\end{equation}
Together, Eqs.~(\ref{internalenergy}) and~(\ref{eq:N}) provide closed-form expressions for the internal energy and particle number of the quon gas in terms of the generalized polylogarithm function $\bar{\text{Li}}_{\nu}(q,\nu;z)$, and thereby fully characterize its thermodynamic behavior.
{\color{black}
We note that the integral in Eq.~\eqref{eq:N}, whose density of states $\Omega(\epsilon)\propto\epsilon^{\nu-1}$ assigns zero weight to $\epsilon=0$, counts only the particles in the excited states, $N_{\rm ex}$. 
The ground-state occupation $\langle n_0\rangle$, given by Eq.~\eqref{Eq:nno}, must be added
separately, $N=N_0+N_{\rm ex}$ with $N_0\equiv\langle n_0\rangle$.}
These two quantities constitute the essential building blocks for the thermodynamic geometry developed in the following section, where they are used to construct the thermodynamic metric of the system.
On this basis, we are subsequently able to investigate how the deformation parameter $q$ affects the geometric properties of the quon gas, establishing a direct link between the generalized quon statistics and the resulting thermodynamic geometry.
\section{Thermodynamic geometry}\label{4}
The geometric formulation of equilibrium thermodynamics, initially pioneered by Weinhold \cite{weinhold1975metric} and Ruppeiner \cite{ruppeiner1979thermodynamics}, establishes an intrinsic Riemannian framework where the thermodynamic parameter space is endowed with a metric structure.
 In Weinhold's formalism, the metric components are derived as the Hessian of the internal energy with respect to extensive thermodynamic variables (such as entropy, volume, and particle number).
  Conversely, Ruppeiner proposed a metric based on the negative second derivatives of the entropy, rooted fundamentally in the Gaussian theory of thermodynamic fluctuations. Although defined on differing thermodynamic potentials, these two formulations are known to be conformally equivalent via temperature rescaling \cite{mrugala1984equivalence}.

Through successive Legendre transformations of the underlying thermodynamic potentials, this geometric approach naturally interfaces with information geometry. Within the grand canonical ensemble, the Fisher--Rao metric tensor is equivalently characterized by the second-order derivatives of the logarithm of the grand partition function $\mathcal{Z}$ with respect to the intensive thermodynamic parameters \cite{janyszek1990riemannian,ruppeiner1995riemannian,crooks2007measuring}:
\begin{equation}\label{metric}
	g_{ij} = \partial_{i}\partial_{j} \ln \mathcal{Z},
\end{equation}
where $\partial_{i}$ denotes the partial derivative with respect to the $i$-th intensive coordinate. For an ideal classical or quantum gas, $\ln \mathcal{Z}$ is an extensive function of the spatial volume $V$, the inverse temperature $\beta = (k_{B}T)^{-1}$, and the dimensionless chemical affinity parameter $\gamma = -\mu / k_{B}T = -\beta\mu$. Under the conventional condition of a fixed spatial volume $V$, the independent thermodynamic state space of the system reduces to a two-dimensional Riemannian manifold parametrized by the coordinate chart $\{\beta, \gamma\}$. Choosing $\gamma$ as the conjugate intensive parameter substantially simplifies both the metric elements and the subsequent curvature computations.

Over the past decades, {\color{black}thermodynamic geometry} and the associated scalar curvature have found widespread and profound applications across diverse areas of modern physics. In classical and interacting systems, the curvature scalar $R$ provides a quantitative probe of microscopic interactions: it vanishes identically for non-interacting classical ideal gases, becomes positive for systems dominated by attractive interactions, and turns negative in the presence of effectively repulsive forces \cite{ruppeiner1995riemannian,janyszek1990riemannian}. The formalism has been extensively exploited to probe the critical behavior and correlation lengths of real fluids, van der Waals systems, and magnetic materials near phase transitions \cite{ruppeiner1995riemannian,ruppeiner2021thermodynamic,ruppeiner2020thermodynamicvdW,lopez2022square}. In quantum contexts, thermodynamic geometry has provided deep insights into systems governed by generalized and deformed statistics, including Tsallis non-extensive statistics \cite{adli2019nonperturbative}, fractional exclusion statistics and trapped gases \cite{mirza2010thermodynamic,ebadi2022thermodynamic,ardabili2026thermodynamic}, deformed quantum algebras \cite{mirza2011thermodynamic,mohammadzadeh2017thermodynamic,esmaili2024thermodynamic,mohammadzadeh2026thermodynamic} and some other generalized statistics \cite{yahyayi2026thermodynamic,seifi2025intrinsic,seifi2025mittag}.
 Concurrently, this geometric perspective has emerged as an {\color{black}important tool} in gravitational physics, particularly in probing the microstructures and phase transitions of black holes in anti-de Sitter spacetimes \cite{mohammadzadeh2021thermodynamic,babaei2022thermodynamic,keshavarzi2025thermodynamic,PhysRevD.100.124033,wei2020extended,sokoliuk2024ads,naveena2021ruppeiner}, where the divergence of $R$ consistently signals the underlying critical and holographic phenomena.

The local and global geometry of this thermodynamic manifold is systematically characterized by the affine connection coefficients (Christoffel symbols of the second kind), which are determined entirely by the metric tensor $g_{ij}$:
\begin{equation}\label{christofel}
	\Gamma^{i}_{jk} = \frac{1}{2} g^{im} \left( \partial_k g_{mj} + \partial_j g_{mk} - \partial_m g_{jk} \right),
\end{equation}
where $g^{mn}$ denotes the components of the inverse metric tensor ($g^{im} g_{mj} = \delta^i_j$), and summation over repeated indices is implicitly assumed. The components of the Riemann curvature tensor describing the intrinsic curvature of the manifold are given by:
\begin{equation}
	R^{i}_{\phantom{i}jkl} = \partial_{k}\Gamma^{i}_{lj} - \partial_{l}\Gamma^{i}_{kj} + \Gamma^{i}_{km}\Gamma^{m}_{lj} - \Gamma^{i}_{lm}\Gamma^{m}_{kj}.
\end{equation}
Contracting the Riemann tensor yields the symmetric second-rank Ricci curvature tensor:
\begin{equation}
	R_{ij} = R^{m}_{\phantom{m}imj}.
\end{equation}
A subsequent trace contraction with the inverse metric tensor produces the Ricci curvature invariant, commonly referred to as the thermodynamic curvature scalar:
\begin{equation}\label{curvature}
	R = g^{ij} R_{ij}.
\end{equation}
For a two-dimensional Riemannian manifold spanned by $\{\beta, \gamma\}$, the calculation of the scalar curvature can be expressed directly in terms of determinants of the metric components and their first partial derivatives \cite{janyszek1990riemannian}:
\begin{equation}\label{thermodynamiccurvature}
	R = -\frac{1}{2\,g^2}
	\begin{vmatrix} 
		g_{\beta\beta} & g_{\beta\gamma} & g_{\gamma\gamma} \\ 
		\partial_\beta g_{\beta\beta} & \partial_\beta g_{\beta\gamma} & \partial_\beta g_{\gamma\gamma} \\ 
		\partial_\gamma g_{\beta\beta} & \partial_\gamma g_{\beta\gamma} & \partial_\gamma g_{\gamma\gamma} 
	\end{vmatrix},
\end{equation}
where $g = \det(g_{ij}) = g_{\beta\beta}g_{\gamma\gamma} - (g_{\beta\gamma})^2$ denotes the metric determinant, and the shorthand comma notation $g_{ij,k} \equiv \partial_k g_{ij}$ has been made explicit.

Having established the foundational geometric framework and the associated curvature invariants, we are now poised to implement this machinery for the quon gas. In the following section, we construct the thermodynamic manifold for the $q$-deformed system, explicitly examine the resulting state-space metric, and assess how the deformation parameter $q$ modulates both the thermodynamic curvature scalar and its critical behavior.

\section{Thermodynamic geometry of quons}\label{5}
In the previous section, we have established the fundamental geometric framework and introduced key curvature invariants relevant to thermodynamic systems.
Building upon this framework, we now construct the parameter space relevant to the thermodynamic description of an ideal quon gas and identify the physically admissible domain of its fugacity.
Throughout this analysis, the \begin{widetext} system volume is kept fixed.
With this constraint, the thermodynamic state space is two-dimensional and can be parametrized by the variables $\beta$ and $\gamma$.
For the subsequent analysis, we introduce the fugacity $z=e^{-\gamma}$ and define the two auxiliary variables $z_{1}=qz$ and $z_{2}=q^{-1}z$.
Using the chain rule and the derivative properties of the polylogarithm functions, the derivatives with respect to $\gamma$ can be written as
\begin{equation}
	\begin{aligned}
		\frac{\partial}{\partial\gamma}\text{Li}^{}_{\nu}(z_{1})=&
		-qz\frac{\partial}{\partial z_{1}}\text{Li}^{}_{\nu}(z_{1})=-\text{Li}^{}_{\nu-1}(z_{1})
		\\	\frac{\partial}{\partial\gamma}\text{Li}^{}_{\nu}(z_{2})=&
		-q^{-1}z\frac{\partial}{\partial z_{2}}\text{Li}^{}_{\nu}(z_{2})=-\text{Li}^{}_{\nu-1}(z_{2}).
		\label{guide}
	\end{aligned}
\end{equation}
These relations lead directly to the corresponding derivative of the generalized polylogarithm,
\begin{equation}\label{dG}
	\frac{\partial}{\partial\gamma}\bar{\text{Li}}^{}_{\nu}(q,\nu;z)=-\bar{\text{Li}}^{}_{\nu-1}(q,\nu;z).
\end{equation}
Using Eqs.~(\ref{metric}), (\ref{internalenergy}), and (\ref{dG}), the components of the thermodynamic metric are obtained as
	\begin{equation}
		\begin{aligned}
			g^{}_{\beta\beta}&=\frac{{{\partial ^2}\ln{\cal{Z}}}}{{\partial {\beta ^2}}} =-(\frac{\partial U}{\partial \beta})^{}_{\gamma}
			=-\frac{\partial}{\partial \beta}\left[\nu{\beta^{-(\nu+1)}}\bar{\text{Li}}^{}_{\nu+1}(q,\nu;z)\right]
			=\frac{\Gamma(\nu+2)}{\Gamma(\nu)}{\beta^{-(\nu+2)}}\bar{\text{Li}}^{}_{\nu+1}(q,\nu;z),
			\\
			g^{}_{\beta \gamma } &= g^{}_{\gamma \beta } =\frac{{{\partial ^2}\ln{\cal{Z}}}}{{\partial {\beta}\partial {\gamma}}}=-(\frac{\partial N}{\partial\beta})^{}_{\gamma}=-\frac{\partial}{\partial\beta}\left[{\beta^{-\nu}}\bar{\text{Li}}^{}_{\nu}(q,\nu;z)\right]
			=\frac{\Gamma(\nu+1)}{\Gamma(\nu)}{\beta^{-(\nu+1)}}\bar{\text{Li}}^{}_{\nu}(q,\nu;z),
			\\
			g^{}_{\gamma \gamma } &= \frac{{{\partial ^2}\ln{\cal{Z}}}}{{\partial {\gamma ^2}}} =-(\frac{\partial N}{\partial\gamma})^{}_{\beta}=-\frac{\partial}{\partial\gamma}\left[{\beta^{-\nu}}\bar{\text{Li}}^{}_{\nu}(q,\nu;z)\right]
			={\beta^{-\nu}}\bar{\text{Li}}^{}_{\nu-1}(q,\nu;z).
			\label{elements}
		\end{aligned}
	\end{equation}
	The derivatives of the metric tensor required for the evaluation of the
	thermodynamic curvature are obtained as :
	\begin{equation}\label{gBBB}
		\begin{aligned}
			&g_{\beta\beta,\beta}=\frac{\partial}{\partial\beta}g_{\beta\beta}=-\frac{\Gamma(\nu+3)}{\Gamma(\nu)}{\beta^{-(\nu+3)}}\bar{\text{Li}}^{}_{\nu+1}(q,\nu;z),
			\\
			&g_{\beta\beta,\gamma}=g_{\beta\gamma,\beta}=g_{\gamma\beta,\beta}=\frac{\partial}{\partial\gamma}g_{\beta\beta}=-\frac{\Gamma(\nu+2)}{\Gamma(\nu)}{\beta^{-(\nu+2)}}\bar{\text{Li}}^{}_{\nu}(q,\nu;z),
			\\
			&g_{\beta\gamma,\gamma}=g_{\gamma\beta,\gamma}=g_{\gamma\gamma,\beta}=\frac{\partial}{\partial\beta}g_{\gamma\gamma}=-\frac{\Gamma(\nu+1)}{\Gamma(\nu)}{\beta^{-(\nu+1)}}\bar{\text{Li}}^{}_{\nu-1}(q,\nu;z),
			\\
			&g_{\gamma\gamma,\gamma}=\frac{\partial}{\partial\gamma}g_{\gamma\gamma}=-{\beta^{-\nu}}\bar{\text{Li}}^{}_{\nu-2}(q,\nu;z).
		\end{aligned}
	\end{equation}
	Substituting Eqs.~(\ref{elements}) and (\ref{gBBB}) into
	Eq.~(\ref{thermodynamiccurvature}), the thermodynamic curvature of the
	quon gas is obtained as
	\begin{equation}
	\begin{aligned}
		{\color{black}R=R(z,q;\nu)}=\frac{\beta^{\nu }(\nu +1)}{2}\Bigg[\frac{\left(\bar{\text{Li}}^{}_{\nu-1}(q,\nu;z)\bar{\text{Li}}^{2}_{\nu}(q,\nu;z)-2\bar{\text{Li}}^{2}_{\nu-1}(q,\nu;z)\bar{\text{Li}}^{}_{\nu+1}(q,\nu;z)\right)}{\left(\nu\bar{\text{Li}}^{2}_{\nu}(q,\nu;z)-(\nu+1)\bar{\text{Li}}^{}_{\nu-1}(q,\nu;z)\bar{\text{Li}}^{}_{\nu+1}(q,\nu;z)\right){}^2}&
		\\+\frac{\bar{\text{Li}}^{}_{\nu-2}(q,\nu;z)\bar{\text{Li}}^{}_{\nu}(q,\nu;z)\bar{\text{Li}}^{}_{\nu+1}(q,\nu;z)}{\left(\nu\bar{\text{Li}}^{2}_{\nu}(q,\nu;z)-(\nu+1)\bar{\text{Li}}^{}_{\nu-1}(q,\nu;z)\bar{\text{Li}}^{}_{\nu+1}(q,\nu;z)\right){}^2}\Bigg]&.
		\label{R}
	\end{aligned}
\end{equation}
\end{widetext}
The thermodynamic curvature of the quon gas is expressed in terms of generalized polylogarithm functions (Eq.~\eqref{R}). In the numerical calculations we set $A = 1$, but $A$ is in fact a dimensionful constant with $[A^{D}] = E^{-\nu}$. Restoring it rescales the curvature as $R \to R/A^{D}$, which compensates the factor $\beta^{\nu}$ appearing in the expression for $R$. The curvature is therefore dimensionless, and the choice $A = 1$ does not affect any of our results. Eq.~(\ref{R}) is used below to examine the dependence of the curvature on the fugacity and the deformation parameter $q$.

\begin{figure}[H]
	\centering
	\begin{subfigure}[b]{0.49\columnwidth}
		\centering
		\includegraphics[width=\textwidth]{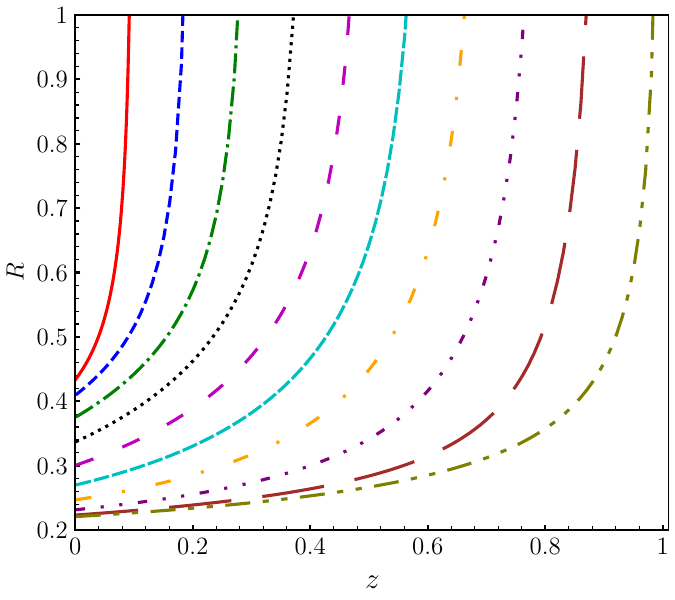}
		\caption{}
		\label{fig:1}
	\end{subfigure}
	\hfill
	\begin{subfigure}[b]{0.49\columnwidth}
		\centering
		\includegraphics[width=\textwidth]{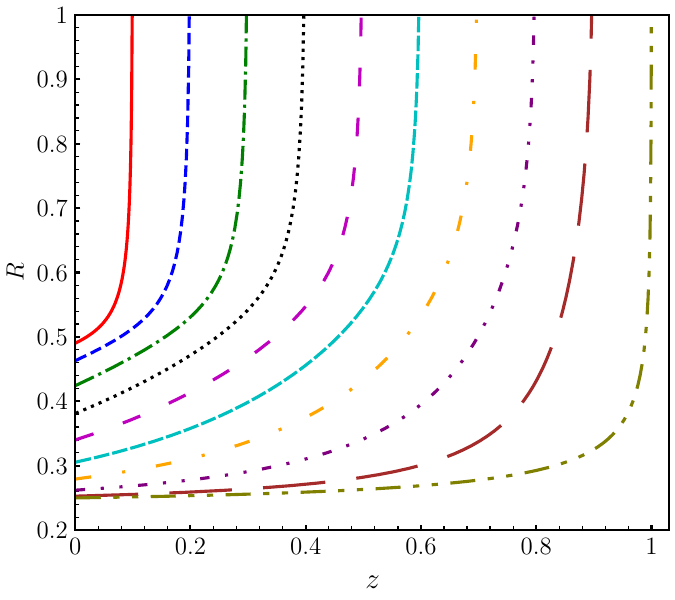}
		\caption{}
		\label{fig:2}
	\end{subfigure}
	\vspace{0.15cm}
	\begin{subfigure}[b]{0.49\columnwidth}
		\centering
		\includegraphics[width=\textwidth]{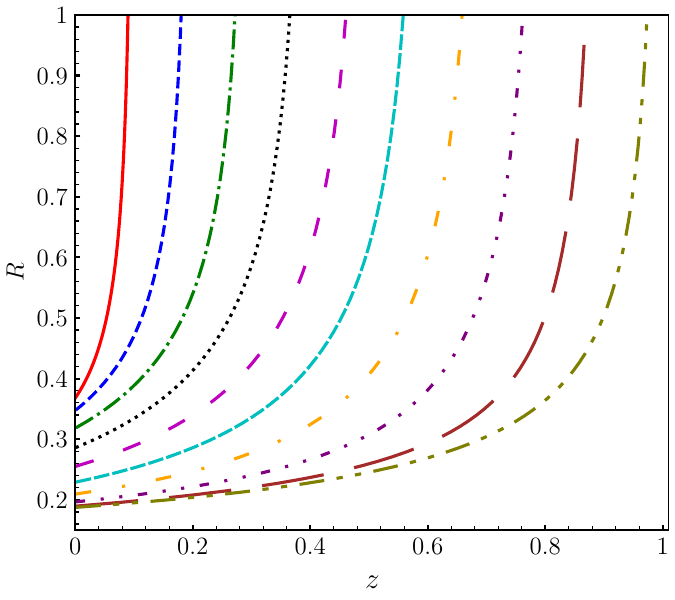}
		\caption{}
		\label{fig:3} 
	\end{subfigure}
	\hfill
	\begin{subfigure}[b]{0.49\columnwidth}
		\centering
		\includegraphics[width=\textwidth]{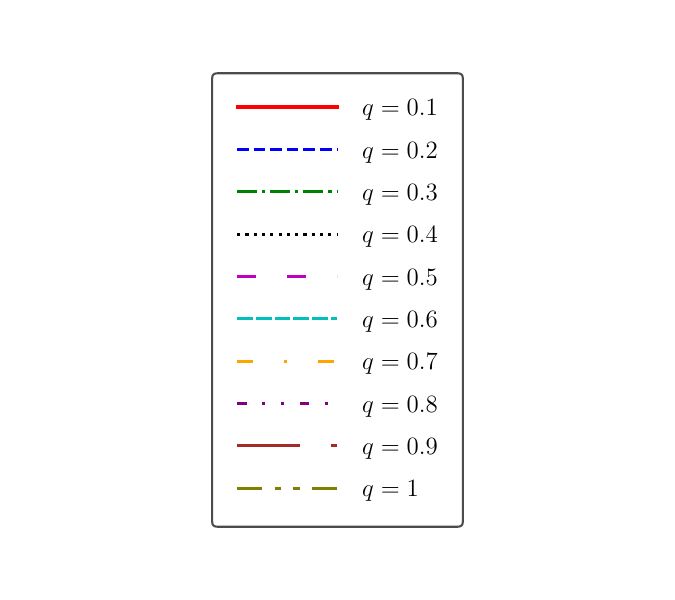}
		\caption{}
		\label{fig:legend} 
	\end{subfigure}
	\caption{\justifying
		Thermodynamic curvature $R$ as a function of the fugacity $z$ for an ideal quon gas with $\beta=1$ : (a) $D=3$, $\sigma=2$; {\color{black}
			(b) $D=2$, $\sigma=2$; (c) $D=2$, $\sigma=1$. 
			(d) The curves in each panel are shown for $q=0.1$ to $q=1$ respectively}.}
		\label{R-Z}
   \end{figure}
Fig.~\ref{R-Z} shows the thermodynamic scalar curvature $R$ as a function of the fugacity $z$ for several representative values of the deformation parameter $q$. \textcolor{black}{To make the dependence on the spatial dimensionality explicit, we plot $R(z)$ for three different choices of $\nu$: (a)  ($\nu=3/2$), (b)  ($\nu=1$), and (c)  ($\nu=2$).
	 As Eq.~\eqref{R} shows, $R$ depends on the combination $\nu=D/\sigma$, so that $R=R(\nu,q,z)=R(D,\sigma,q,z)$. The different curves demonstrate that changing $D$ and $\sigma$ modifies the magnitude and steepness of the divergence of $R$ near $z=1$, while the overall qualitative behavior (sign structure and divergence at the critical point) remains the same for all cases. 
}
\section{Physically Admissible Fugacity Domain and Thermodynamic Geometry in the Condensed Phase}\label{6}
The requirement that the average occupation number of the ground state remain positive, $\langle n_0 \rangle \geq 0$, leads to two mathematically disconnected fugacity intervals for $0 < q < 1$: the lower interval $0 < z < q$ given by Eq.~(\ref{part1}) and the upper interval {\color{black}$(1+q_{}^2) /2q< z < 1/q$} given by Eq.~(\ref{part2}). These two branches are separated by an intermediate gap where the occupation number becomes negative. Although the upper interval algebraically satisfies the positivity condition, determining its physical validity requires examining the global thermodynamic behavior of the system as a function of temperature.

\begin{figure}[H]
	\centering
	\begin{subfigure}[b]{0.48\columnwidth}
		\centering
		\includegraphics[width=\textwidth]{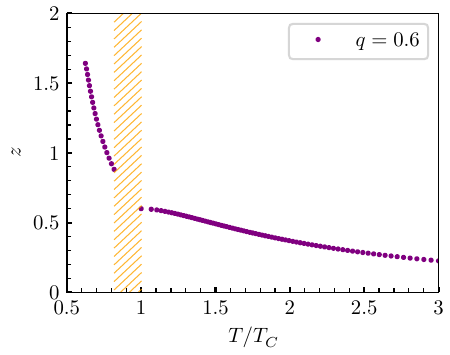}
		\caption{}
		\label{fig:z_full}
	\end{subfigure}
	\hfill
	\begin{subfigure}[b]{0.48\columnwidth}
		\centering
		\includegraphics[width=\textwidth]{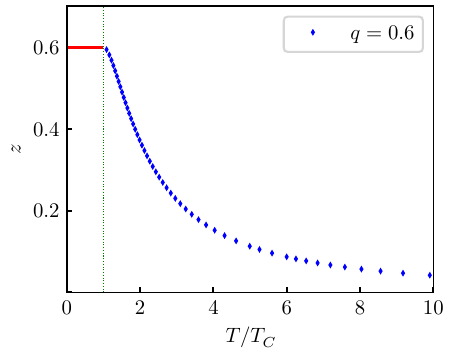}
		\caption{}
		\label{fig:z_phys}
	\end{subfigure}
	\vspace{0.15cm}
	\begin{subfigure}[b]{0.48\columnwidth}
		\centering
		\includegraphics[width=\textwidth]{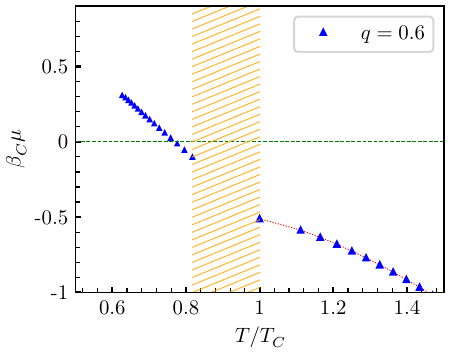}
		\caption{}
		\label{fig:mu_full}
	\end{subfigure}
	\hfill
	\begin{subfigure}[b]{0.48\columnwidth}
		\centering
		\includegraphics[width=\textwidth]{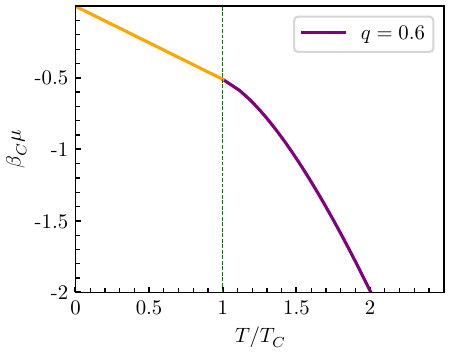}
		\caption{}
		\label{fig:mu_phys}
	\end{subfigure}
	\caption{\justifying
		Fugacity $z$ and reduced chemical potential $\mu/(k_B T_c)$ as functions of the reduced temperature $T/T_c$ for $q=0.6$. Panels (a) and (c) display the complete mathematical solutions, illustrating the thermodynamic discontinuity and the inaccessible branch. Panels (b) and (d) present the physically admissible branch retained in our framework.}
	\label{fig:z_compare}
\end{figure}

To assess whether this second branch represents physically realizable thermodynamic states, we investigate the temperature dependence of the fugacity and the chemical potential. Evaluating the total particle-number equation, Eq.~(\ref{eq:N}), at the critical condensation temperature $T_c$, where the fugacity reaches the threshold $z_c = q$, yields the relation for the reduced temperature:
\begin{equation}
	\frac{T}{T_c}
	=
	\left(
	\frac{\mathrm{Li}_{\nu}(qz)+\mathrm{Li}_{\nu}(z/q)}
	{\mathrm{Li}_{\nu}(q^2)+\mathrm{Li}_{\nu}(1)}
	\right)^{-\frac{1}{\nu}}.
	\label{eq:T_Tc_z}
\end{equation}
Along the second branch, where {\color{black}$z > (1+q^2)/2q$}, the argument of the second polylogarithm exceeds unity ($z/q > 1$).  
In this domain, the standard series expansion of the polylogarithm diverges;
	however, its integral representation \textcolor{black}{can still be evaluated unambiguously using the Cauchy principal value prescription, whose details are presented in the Appendix of Ref.~\cite{yahyayi2026thermodynamic}.}
 Combining Eq.~(\ref{eq:T_Tc_z}) with the fundamental relation $\mu / (k_B T_c) = (T/T_c)\ln z$, the complete mathematical solutions for $z(T/T_c)$ and $\mu(T/T_c)$ are obtained and displayed in Figs.~\ref{fig:z_compare}(a) and \ref{fig:z_compare}(c) for $q=0.6$.

As clearly demonstrated in Figs.~\ref{fig:z_compare}(a) and \ref{fig:z_compare}(c), the emergence of the second branch introduces severe thermodynamic inconsistencies. First, it is completely disconnected from the high-temperature classical dilute limit where $z \to 0$ as $T \to \infty$. Second, it produces an unphysical gap accompanied by an abrupt discontinuity in both the fugacity and chemical potential curves across $T_c$. Third, it renders a portion of the temperature regime below $T_c$ thermodynamically inaccessible and multi-valued.

Imposing the fundamental requirement of thermodynamic continuity and single-valuedness dictates that the second branch must be excluded from the physical state space.
 Consequently, the physically admissible fugacity domain is uniquely restricted to the continuous lower branch:
\begin{equation}
	0 < z \le q,
\end{equation}
with the maximal physical fugacity given by
\begin{equation}
	z_{\mathrm{max}} = z_c = q.
\end{equation}
The physically admissible trajectories for $z$ and $\mu/(k_B T_c)$ are presented in Figs.~\ref{fig:z_compare}(b) and \ref{fig:z_compare}(d).

Along the physical branch, as the temperature decreases from the high-temperature
{\color{black} regime ($z\to0$ as $T\to\infty$) toward $T_c$, the fugacity increases monotonically
and continuously until it reaches saturation at $z_c=q$. This mirrors the ideal Bose
gas ($q\to1$ in Eq.~\eqref{eq:N}), where the particle-number equation
$N=\beta^{-\nu}\mathrm{Li}_\nu(z)$ admits a solution with $z\le1$ only for $T\ge T_c$;
below $T_c$ the excess particles occupy the ground state and the fugacity remains
pinned at its maximal allowable value. In the same way, for the quon gas
Eq.~\eqref{eq:N} can be solved for $z(T)\le q$ only down to $T_c$, and for $T\le T_c$ the
fugacity remains pinned at}
\begin{equation}\label{Eq:ztq}
	z(T) = q, \quad \text{for } T \le T_c.
\end{equation}
{\color{black}
By contrast, if the second branch were admitted, Eq.~\eqref{eq:T_Tc_z} would yield solutions
$z(T/T_c)$ disconnected from this continuous trajectory, leaving a gap of
temperatures that cannot be reached by any continuous thermodynamic path
[Figs.~\ref{fig:z_compare}(a) and \ref{fig:z_compare}(c)]. 

Consequently, a range of temperatures below the condensation temperature is inaccessible via this branch; therefore, this branch cannot represent a physical state of the system
, and the region $T\le T_c$ is described solely by the saturated,
temperature-independent fugacity $z=q$.}

{\color{black}
The fraction of particles in the ground state, which lies outside the
continuous integration of Eq.~\eqref{eq:N}, follows from $N=N_0+N_{\rm ex}$. In the
normal phase ($z<q$), $\langle n_0\rangle$ is finite and $N_0/N=O(1/N)$
vanishes in the thermodynamic limit. For $T\le T_c$ ($z=q$), one has
$N_{\rm ex}=(k_BT)^{\nu}\,\bar{\rm Li}_\nu(q,\nu;q)$ and
$N=(k_BT_c)^{\nu}\,\bar{\rm Li}_\nu(q,\nu;q)$, so that
\begin{equation}
	\frac{N_0}{N}=1-\left(\frac{T}{T_c}\right)^{\nu},\qquad T\le T_c,
	\label{eq:N0}
\end{equation}
which has the same form as for the ideal Bose gas, while $T_c$ is shifted by
the deformation (Eq.~\eqref{eq:T_Tc_z}); for $\nu=3/2$, $T_c(q)/T_c(1)\simeq1.57,\,1.44,\,1.22$
for $q=0.2,\,0.6,\,0.9$. For finite $N$, the fugacity does not sit exactly at
$q$: writing $z=q\,e^{-\varepsilon}$, only the second term of Eq.~\eqref{Eq:nno} is
singular, $\langle n_0\rangle\simeq 1/(2\varepsilon)$, which gives
\begin{equation}
	z\simeq q\left(1-\frac{1}{2N_0}\right),
	\label{eq:zN0}
\end{equation}
the quon counterpart of $z\simeq1-1/N_0$ for the Bose gas. The pinning in
Eq.~\eqref{Eq:ztq} is thus the $N_0\to\infty$ limit of this relation.}
Despite this analogy, a distinctive feature of quon statistics emerges in the behavior of the chemical potential. While in an ideal Bose gas the pinning $z=1$ forces the chemical potential to vanish identically ($\mu = 0$) throughout the condensed phase, for the quon gas the chemical potential retains an explicit linear dependence on temperature:
\begin{equation}
	\mu(T) = k_B T \ln q, \quad \text{for } T \le T_c.
\end{equation}
At the condensation threshold, this yields the critical value $\mu_c / (k_B T_c) = \ln q$, recovering the standard vanishing bosonic potential only in the limit $q \to 1$.
\begin{figure}[t]
	\centering
	\begin{subfigure}[b]{0.48\columnwidth}
		\centering
		\includegraphics[width=\textwidth]{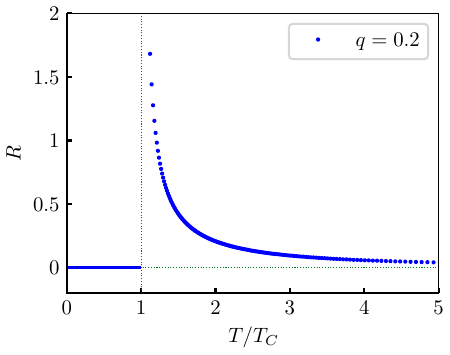}
		\caption{}
		\label{fig:R_q02}
	\end{subfigure}
	\hfill
	\begin{subfigure}[b]{0.48\columnwidth}
		\centering
		\includegraphics[width=\textwidth]{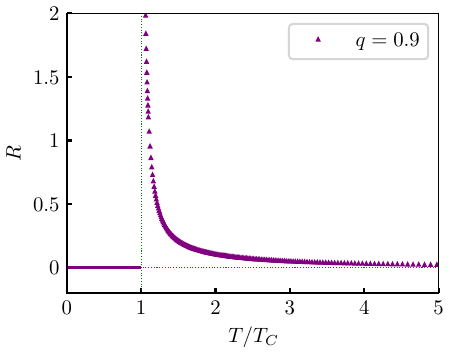}
		\caption{}
		\label{fig:R_q09}
	\end{subfigure}
	\caption{\justifying
		Thermodynamic scalar curvature $R$ as a function of the reduced temperature $T/T_c$ for $q=0.2$ (panel a) and $q=0.9$ (panel b).}
	\label{fig:R_vs_T_final}
\end{figure}
With the physical temperature dependence of fugacity established across all temperature regimes, we examine the thermodynamic scalar curvature $R(T/T_c)$, shown in Fig.~\ref{fig:R_vs_T_final} for $q=0.2$ and $q=0.9$. 

\begin{figure}[t]
	\centering
	\includegraphics[width=0.98\columnwidth]{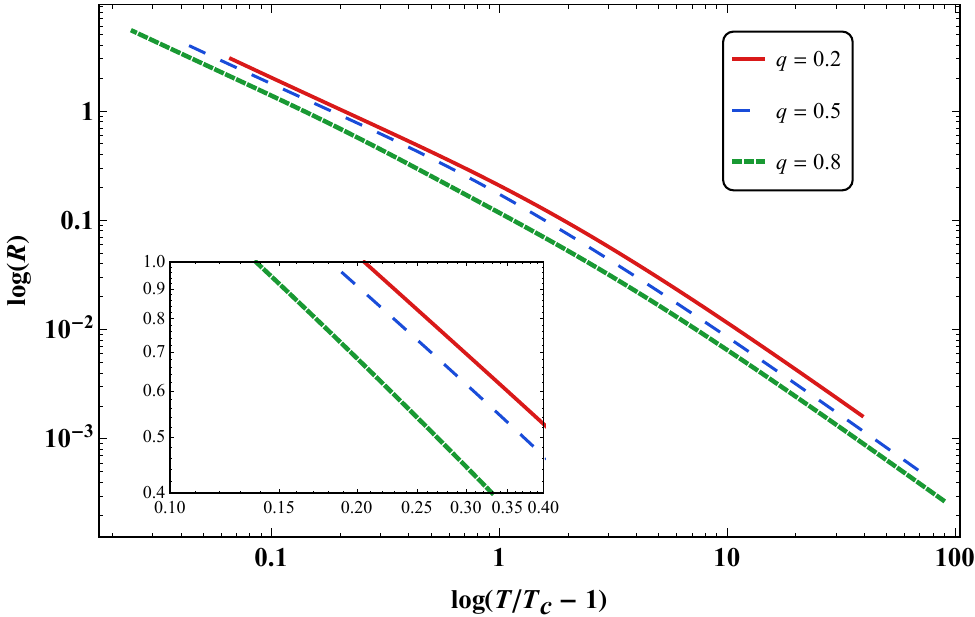}
	\caption{{\color{black}
	A log--log plot of the thermodynamic curvature $R$ versus the reduced
	temperature $t=T/T_c-1$ above the critical point, for $\nu=3/2$ and
	$q=0.2,\,0.5,\,0.8$ the dashed lines have slopes of $-0.99859\pm 0.00004, -0.99869\pm 0.00004,$ and $-0.99850\pm 0.00004$ respectively, indicating that $R \sim t^{-1}$.
	The deformation parameter shifts the curves vertically (amplitude) but does
	not change the slope.}
}
	\label{fig:Rexp}
\end{figure}

In the normal phase ($T > T_c$), the curvature $R$ increases sharply and diverges as $T \to T_c^+$ ($z \to q$). In {\color{black}thermodynamic geometry}, this divergence serves as {\color{black}a clear geometric signature} of macroscopic correlation growth and critical fluctuations near the condensation transition.
{\color{black}
To characterize this divergence quantitatively, note that near $T_c$ only
$\mathrm{Li}_s(z/q)$ in $\bar{\rm Li}_s(q,\nu;z)$ is singular, since
$\mathrm{Li}_s(e^{-\varepsilon})\simeq\Gamma(1-s)\,\varepsilon^{s-1}$ with
$z=q\,e^{-\varepsilon}$. For $\nu=3/2$, Eq.~\eqref{R} then gives
$R\sim\varepsilon^{-1/2}$, and Eq.~\eqref{eq:T_Tc_z} at fixed $N$ gives
$\varepsilon\propto t^{2}$ with $t=T/T_c-1$, so that
\begin{equation}
	R \sim t^{-1}, \qquad T\to T_c^{+},
	\label{eq:Rexp}
\end{equation}
in agreement with the numerical results of Fig.~\ref{fig:Rexp}. This is the
same exponent as in the ideal Bose gas: the deformation $q$ only rescales the
amplitude and leaves the critical behavior unchanged.}

In the condensed phase ($T \le T_c$), the saturation condition $z = q = \text{const.}$ fixes the fugacity, thereby eliminating one independent thermodynamic degree of freedom. Consequently, the two-dimensional fluctuating state space collapses into an effectively one-dimensional manifold. Because the Riemann curvature tensor vanishes identically for any one-dimensional Riemannian manifold, the thermodynamic scalar curvature vanishes throughout the condensed regime:
\begin{equation}
	R = 0, \quad \text{for } T \le T_c.
\end{equation}
\textcolor{black}{
	The vanishing of $R$ for $T\le T_c$ is a geometric consequence of the dimensional reduction of the state space and should not be read as the absence of thermodynamic correlations; energy fluctuations remain nonzero, $g^{}_{\beta\beta}=k^{}_{\mathrm B} T^2_{}C^{}_V>0$.}
	{\color{black}
	Note that Eqs.~\eqref{thermodynamiccurvature} and \eqref{R} are valid only for a two-dimensional state space; with the loss of one degree of freedom at $z=q$, the curvature in the condensed phase does not follow Eq.~\eqref{thermodynamiccurvature}, and the vanishing of $R$ is instead a property of the resulting one-dimensional manifold\cite{DIAZSALDANA2024416042}, as proved in Appendix~\ref{app:ricci_1d}. Consequently, the limit $z\to q$ of Eq.~\eqref{R} (the divergence of $R$ at $T_c^+$) and the value $R=0$ for $T\le T_c$ are not equivalent.}
{\color{black}
\subsection{Specific heat and its relation to the scalar curvature}

To place the curvature results in a conventional thermodynamic context, we
compute the isochoric specific heat $C_V=(\partial U/\partial T)_{N,V}$. 

\paragraph{Normal phase, $T>T_c$.} Keeping $N$ fixed gives
$(T/z)\,dz/dT=-\nu \bar{\text{Li}}^{}_{\nu}(z)/\bar{\text{Li}}^{}_{\nu-1}(z)$, and therefore
\begin{equation}
	\frac{C_V}{Nk_{\mathrm B}}
	=\nu(\nu+1)\,\frac{\bar{\text{Li}}^{}_{\nu+1}(z)}{\bar{\text{Li}}^{}_{\nu}(z)}
	-\nu^{2}\,\frac{\bar{\text{Li}}^{}_{\nu}(z)}{\bar{\text{Li}}^{}_{\nu-1}(z)} .
	\label{eq:CV_normal}
\end{equation}
For $q\to1$ this reduces to the ideal Bose-gas result (for $\nu=3/2$:
$\tfrac{15}{4}\text{Li}^{}_{5/2}(z)/\text{Li}^{}_{3/2}(z)-\tfrac{9}{4}\text{Li}^{}_{3/2}(z)/\text{Li}^{}_{1/2}(z)$), and for
$T\to\infty$ ($z\to0$) it tends to the classical value $C_V\to\nu Nk_{\mathrm B}$.

\paragraph{Condensed phase, $T\le T_c$.} Here the fugacity is pinned at $z=q$,
the condensate carries no energy, and $N=k_{\mathrm B}^{\nu}T_c^{\,\nu}\bar{\text{Li}}^{}_{\nu}(q)$.
The second term of Eq.~\eqref{eq:CV_normal} is absent and
\begin{equation}
	\begin{aligned}
	    \frac{C_V}{Nk_{\mathrm B}}
	&=\nu(\nu+1)\,\frac{\bar{\text{Li}}^{}_{\nu+1}(q)}{\bar{\text{Li}}^{}_{\nu}(q)}\left(\frac{T}{T_c}\right)^{\nu},
    \\\bar{\text{Li}}^{}_{s}(q)&\equiv\mathrm{Li}_s(q^{2})+\zeta(s).
	\end{aligned}
	\label{eq:CV_cond}
\end{equation}
Thus $C_V\neq0$ below $T_c$ and scales as $T^{\nu}$, exactly as for the
ordinary Bose gas.

\paragraph{Comparison with the scalar curvature.} For $\nu=3/2$ the divergence
$R\propto(T-T_c)^{-1}$ as $T\to T_c^{+}$ is \emph{not} accompanied by a
divergence of $C_V$, which only displays a cusp. The singularity of $R$ is
instead tied to the diverging number fluctuations, since
$g_{\gamma\gamma}=\langle\Delta N^{2}\rangle\propto \bar{\text{Li}}^{}_{\nu-1}(z)$ diverges
(equivalently, the isothermal compressibility), while
$g_{\beta\beta}\propto \bar{\text{Li}}^{}_{\nu+1}$ remains finite. 
In this sense $R$ probes the
density-fluctuation singularity of the quon gas, whereas the cusp of $C_V$
provides the independent thermal signature of the transition.
Furthermore, $C_V\neq0$ for $T<T_c$ is fully consistent with $R=0$ there:
the vanishing curvature is a geometric consequence of the reduced
one-dimensional state space, whereas the remaining metric component is
$g_{\beta\beta}=k_{\mathrm B}T^{2}C_V>0$, so that energy fluctuations persist
in the condensed phase.
\begin{figure}[t]
	\centering
	\begin{subfigure}[b]{0.48\columnwidth}
		\centering
		\includegraphics[width=\textwidth]{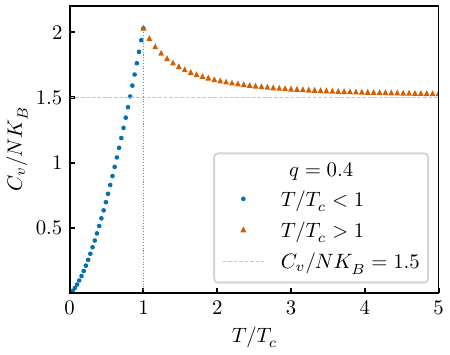}
		\caption{}
		\label{fig:QuonCv04}
	\end{subfigure}
	\hfill
	\begin{subfigure}[b]{0.48\columnwidth}
		\centering
		\includegraphics[width=\textwidth]{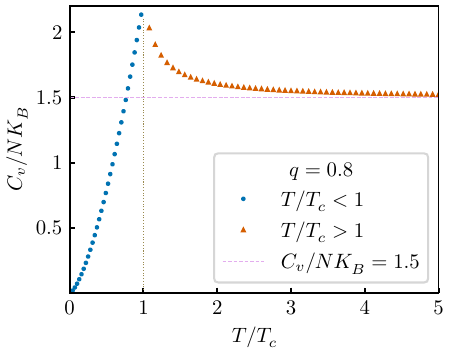}
		\caption{}
		\label{fig:QuonCv08}
	\end{subfigure}
	\caption{\justifying
		Specific heat $C^{}_v/N K^{}_B$ as a function of the reduced temperature $T/T_c$ for $q=0.4$  and $q=0.8$ with $D=3$ and $\sigma=2$.}
	\label{fig:QuonCv}
\end{figure}
}

\section{Conclusion}\label{7}
In this work, we have developed a unified thermodynamic and geometric formulation of an ideal quon gas in the grand canonical ensemble. The introduction of the quon deformation parameter continuously modifies the underlying quantum statistics, providing a fertile ground for exploring how statistical deformations manifest in both macroscopic observables and the intrinsic Riemannian structure of thermodynamic equilibrium states.

A central outcome of our investigation is the resolution of the physical domain of the fugacity. The positivity requirement imposed on the average ground-state occupation number naturally divides the mathematical state space into two disconnected branches separated by an unphysical gap. 
By analyzing the global thermodynamic behavior as a function of temperature; facilitated by evaluating polylogarithm representations via the Cauchy principal value, {\color{black} we have shown that the second branch is not physically admissible within the present formulation.}
 It is decoupled from the classical dilute limit, introduces an unphysical discontinuity at the transition point, and yields an inaccessible, multi-valued temperature regime. Consequently, the physically admissible state space is uniquely restricted to the continuous lower domain, establishing the deformation parameter itself as the critical condensation boundary. This threshold directly governs the critical condensation temperature, giving rise to a generalized condensation mechanism where the fugacity saturates below the critical point, accompanied by an explicit linear temperature dependence of the chemical potential.

To investigate the microscopic statistical correlations geometrically, we constructed the Fisher-Rao thermodynamic metric on the equilibrium state manifold and derived the corresponding scalar curvature. Throughout the entire physical domain, the curvature remains strictly positive for all admissible deformation parameters. In the language of {\color{black}thermodynamic geometry}, this persistent positivity  signifies that quon statistics preserves the intrinsically attractive statistical behavior characteristic of bosons, exhibiting no crossover into effective statistical repulsion. 

Crucially, the scalar curvature serves as an invariant geometric probe of the condensation transition. As the system approaches the critical boundary from the normal phase, the curvature experiences a steep increase and diverges precisely at the critical threshold. This divergence mirrors the macroscopic growth of correlation volume and critical fluctuations near the transition, demonstrating that the boundary of the accessible thermodynamic state space is intrinsically singular. In contrast, within the condensed phase below the transition temperature, the saturation of fugacity effectively freezes one thermodynamic degree of freedom. This dimensional reduction collapses the fluctuating Riemannian manifold, causing the thermodynamic scalar curvature to vanish identically throughout the condensed regime.

\bibliography{refs}
\appendix

{\color{black}
	\section{Vanishing of Ricci Curvature in One Dimension}
\label{app:ricci_1d}

In this appendix, we explicitly demonstrate that the Ricci curvature tensor vanishes identically for any one-dimensional Riemannian manifold (curve) parametrized by a coordinate $x$.

Let the line element of a one-dimensional manifold be given by:
\begin{equation}
    ds^2 = g_{11}(x) \, dx^2,
\end{equation}
where $g_{11}(x) > 0$ is a smooth scalar metric function. The inverse metric component is simply $g^{11} = (g_{11})^{-1}$.

The non-vanishing Christoffel symbol of the Levi-Civita connection, $\Gamma^1_{11}$, is defined as:
\begin{equation}
    \Gamma^1_{11} = \frac{1}{2} g^{11} \left( \partial_1 g_{11} + \partial_1 g_{11} - \partial_1 g_{11} \right) = \frac{1}{2g_{11}} \frac{dg_{11}}{dx}.
\end{equation}

The Riemann curvature tensor in index notation is expressed as:
\begin{equation}
    R^\rho_{\ \sigma\mu\nu} = \partial_\mu \Gamma^\rho_{\nu\sigma} - \partial_\nu \Gamma^\rho_{\mu\sigma} + \Gamma^\rho_{\mu\lambda}\Gamma^\lambda_{\nu\sigma} - \Gamma^\rho_{\nu\lambda}\Gamma^\lambda_{\mu\sigma}.
\end{equation}

For a 1-dimensional manifold, the only independent component is $R^1_{\ 111}$. Substituting $\mu = \nu = 1$:
\begin{equation}
    R^1_{\ 111} = \partial_1 \Gamma^1_{11} - \partial_1 \Gamma^1_{11} + (\Gamma^1_{11})^2 - (\Gamma^1_{11})^2 \equiv 0.
\end{equation}
This also follows immediately from the antisymmetry of the Riemann tensor in its last two indices ($R^\rho_{\ \sigma\mu\nu} = -R^\rho_{\ \sigma\nu\mu}$).

The Ricci curvature tensor, defined by the contraction $R_{\mu\nu} = R^\rho_{\ \mu\rho\nu}$, consequently reduces to:
\begin{equation}
    R_{11} = R^1_{\ 111} = 0.
\end{equation}

Therefore, the intrinsic curvature of any one-dimensional space is identically zero, implying that all 1D Riemannian manifolds are locally flat.}

\end{document}